\documentclass[10pt,a4paper]{article}

\usepackage[latin1]{inputenc}
\usepackage{fourier}
\usepackage{graphicx}
\usepackage{amsmath}
\usepackage{amsfonts}
\usepackage[latin1]{inputenc}
\usepackage[T1]{fontenc}
\usepackage{verbatim}
\usepackage{latexsym}
\usepackage{natbib}

\usepackage{hyperref}
\hypersetup{backref,colorlinks=true,citecolor=blue}

\usepackage{alltt}
\usepackage{color}                       
\definecolor{grey}{rgb}{0.9,0.9,0.9}

\usepackage{etoolbox}
\makeatletter
\patchcmd{\NAT@citex}
  {\@citea\NAT@hyper@{%
     \NAT@nmfmt{\NAT@nm}%
     \hyper@natlinkbreak{\NAT@aysep\NAT@spacechar}{\@citeb\@extra@b@citeb}%
     \NAT@date}}
  {\@citea\NAT@nmfmt{\NAT@nm}%
   \NAT@aysep\NAT@spacechar\NAT@hyper@{\NAT@date}}{}{}
\patchcmd{\NAT@citex}
  {\@citea\NAT@hyper@{%
     \NAT@nmfmt{\NAT@nm}%
     \hyper@natlinkbreak{\NAT@spacechar\NAT@@open\if*#1*\else#1\NAT@spacechar\fi}%
       {\@citeb\@extra@b@citeb}%
     \NAT@date}}
  {\@citea\NAT@nmfmt{\NAT@nm}%
   \NAT@spacechar\NAT@@open\if*#1*\else#1\NAT@spacechar\fi\NAT@hyper@{\NAT@date}}
  {}{}
\makeatother

\begin{document}

\title{On the Coordinate System of Space-Weather HMI Active Region Patches (SHARPs): A Technical Note}

\author{Xudong Sun$^{1}$ for the HMI Team \\
\small $^{1}$ HEPL, Stanford University, Stanford, CA 94305 (\href{mailto:xudong@sun.stanford.edu}{xudong@sun.stanford.edu})}

\date{\normalsize \textit{Sep 2, 2022}}

\maketitle

\begin{abstract}
We describe the coordinate systems of two streams of HMI active region vector data. A distinction is made between (a) the 2D grid on which the field vector is measured (or sampled), and (b) the 3D coordinate established at each grid point, in which the field vector is presented. The HMI data reduction can involve coordinate changes on both, with those performed on the former termed ``\textit{remapping}'', the latter ``\textit{vector transformation}''. Relevant pipeline procedures are described. Useful examples are given for data analysis.
\end{abstract}

\section{The Basics}
\label{sec:basics}

The HMI pipeline currently produces two streams of active region vector data. One is \texttt{hmi.sharp\_720s}, the other \texttt{hmi.sharp\_cea\_720s}. For a quick start on the HMI data retrieval, see the \href{http://jsoc.stanford.edu/Priya/JSOC/How_toget_data.html}{FAQ page} and \href{http://jsoc.stanford.edu/ajax/RecordSetHelp.html}{a few examples}.

The series \texttt{hmi.sharp\_720s} aims to provide the vector data in its original format, i.e. a direct cutout from the full-disk image with no coordinate change or interpolation. This allows more flexibility in data analysis. The series \texttt{hmi.sharp\_cea\_720s} aims to provide an easy-to-use format, and involves further data reduction which we will describe below. For details on their contents, see the \href{http://sun.stanford.edu/~mbobra/spaceweather/sharps.htm}{SHARP description page}.

We make a distinction between two kinds of coordinate systems. The first kind is a 2D grid on which the field is measured. Examples include the ``native coordinate'' defined by the pixel arrays on the CCD detector, and the equally spaced grid in Carrington longitude and sine of latitude, used in synoptic maps. The second kind is a 3D coordinate established at each grid point, in which the field vector is presented. The three base vectors may be that from the Heliocentric Earth Ecliptic (HEE) Cartesian coordinate \citep{thompson2006}, which are identical globally. Or, they may be $(\hat{\bf{e}}_r,\hat{\bf{e}}_\theta,\hat{\bf{e}}_\phi)$ as in a Heliocentric spherical coordinate, where their physical directions change in space.

We call the coordinate transformation on the first kind ``\textit{remapping}''. It is also called ``map projection'' in a sense that a 3D solar surface is ``projected'' on to a 2D image. However, such ``projection'' is strictly a mapping of the loci. It should not be confused with any operations on the field vectors. A variety of remapping methods and the resultant map coordinates are described, for example, in \cite{calabretta2002}.

To report the vector field measurement, one needs to select a set of base vectors at each grid point. The field vectors are then decomposed on to them. We call this step ``\textit{vector transformation}''. There is no reason why this cannot be independent from the remapping, i.e. the base vectors can be arbitrary.

The series \texttt{hmi.sharp\_720s} preserves vector data in the native coordinate, i.e. a 2D array as measured at each CCD pixel. The resultant image portrays the Sun from a single perspective at a large distance, and is defined to be in a Helioprojective-cartesian coordinate \citep{thompson2006}. For a nice introductory read for the World Coordinate System (WCS) convention, see ``\href{http://sun.stanford.edu/~beck/JSOC/HMI_WCS_Dummies.pdf}{WCS for Dummies}''.

The three base vectors here are $(\hat{\bf{e}}_\xi,\hat{\bf{e}}_\eta,\hat{\bf{e}}_\zeta)$, with $+\xi$ referring to $+x$ on the CCD array, $+\eta$ as $+y$, and $+\zeta$ as $+z$ out of the image plane. Note that $+\zeta$ coincides with the line-of-sight (LOS) direction. The field vectors are expressed as field strength ($B$), inclination ($\gamma$) and azimuth ($\psi$). The inclination $\gamma$ is defined with respect to $+\zeta$, with 0$^\circ$ pointing out of the image, 180$^\circ$ into the image. The azimuth $\psi$ is defined with respect to $+\eta$, with 0$^\circ$ at $+\eta$ and increase counterclockwise.

To express the field using the base vector $(\hat{\bf{e}}_\xi,\hat{\bf{e}}_\eta,\hat{\bf{e}}_\zeta)$, one uses
\begin{equation}
\label{eq:bvec}
\begin{split}
B_\xi & = - B \sin\gamma \sin \psi, \\
B_\eta & = B \sin\gamma \cos \psi, \\
B_\zeta & = B \cos\gamma.
\end{split}
\end{equation}
Equation set~\eqref{eq:bvec} may be viewed as a simple vector transformation. It is worthwhile to note that the HMI has a $p$-angle of about $180^\circ$, i.e. the north is close to $-\eta$, and the west is close to $-\xi$.

In the following section, we describe how the series \texttt{hmi.sharp\_cea\_720s} is constructed. Again, a distinction is made between the remapping and the vector transformation.


\section{The CEA Series}
\label{sec:cea}

\subsection{Remapping}
\label{subsec:remapping}

The series \texttt{hmi.sharp\_cea\_720s} uses a Cylindrical equal area projection (CEA). The standard CEA coordinates $(x,y)$ relate to the Heliographic longitude and latitude $(\phi,\lambda)$ (see equations (79) and (80) in \cite{calabretta2002})
\begin{equation}
\label{eq:cea}
\begin{split}
x & = \phi, \\
y & = (180^\circ/\pi) \sin\lambda.
\end{split}
\end{equation}
This is also the coordinate used in most synoptic maps, which assumes a reference latitude of zero, i.e. the equator. Each pixel in this map represents a same area. It is suitable for the computation of total flux and perhaps extrapolation, since no special treatment is needed to account for the pixel sizes.


\begin{figure}[t]
\centerline{\includegraphics{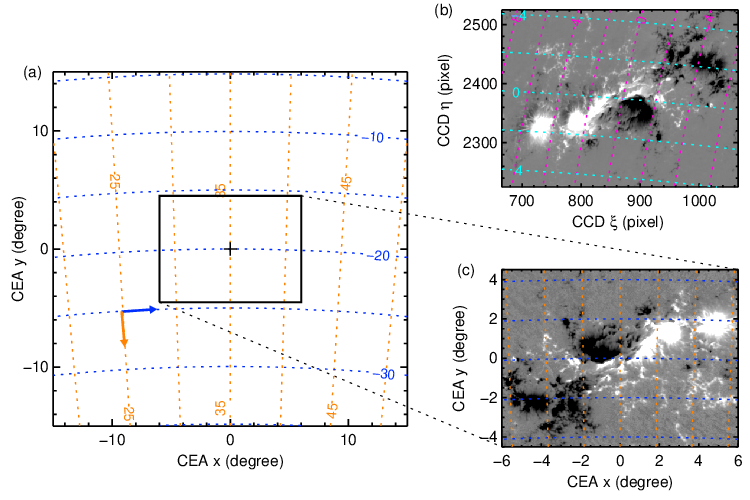}}
\caption{Example of CEA grid and remapping. (a) A $30^\circ\times30^\circ$ large patch. The cross shows the patch center, which corresponds to a Heliographic latitude of $-20^\circ$ and longitude 35$^\circ$. The blue dashed lines are contours of constant Heliographic latitude, as derived from equation set~\eqref{eq:map2sphere}. The orange contours are for longitude. The $12^\circ\times9^\circ$ rectangle shows the field of view for (c). The base vector ${\hat{\bf{e}}}_\phi$ is tangent to the constant latitude contours. It does not align with ${\hat{\bf{e}}}_x$ except at the patch center, as illustrated by the blue arrow. This is similar for ${\hat{\bf{e}}}_\theta$, which points toward south as illustrated by the orange arrow. (b) A $400\times300$ cutout of $B_\zeta$ (line-of-sight field) of AR 11158 at 2011-02-16T18:00Z. The $p$-angle is about 180$^\circ$, so the image is upside down. At W35S20, the foreshortening effect is already obvious. The pink and cyan contours are for constant CEA $x$ and $y$, derived from the inverse of equation sets~\eqref{eq:imgrot} and \eqref{eq:sphere2img}, and equation set~\eqref{eq:sphere2map}. (c) A $12^\circ\times9^\circ$ CEA patch derived from (b) showing the $B_r$ component. Both $B_\zeta$ and $B_r$ are scaled between $\pm$800 G.\label{f:cea}}
\end{figure}


The HMI pipeline uses the patch center Heliographic longitude and latitude $(\phi_c,\lambda_c)$ as the reference pixel. For a set of desired $(x,y)$, the corresponding $(\phi,\lambda)$ are given by
\begin{equation}
\label{eq:map2sphere}
\begin{split}
\lambda & = \sin^{-1} [\, \cos \lambda_c\,y + \sin \lambda_c \, \sqrt{1-y^2} \, \cos x \,], \\
\phi & = \sin^{-1} [\, \sqrt{1-y^2} \, \sin x \, / \cos \lambda \,] + \phi_c.
\end{split}
\end{equation}
Here, $x$ and $y$ have been converted from degree to radian. The patch center coordinates $(x_c,y_c)$ are taken to be $(0,0)$. It is easy to verify that for $\lambda_c=0$, equation set~\eqref{eq:map2sphere} simplifies to the inverse of equation set~\eqref{eq:cea}. It is also easy to verify that the inverse of equation set~\eqref{eq:map2sphere} is
\begin{equation}
\label{eq:sphere2map}
\begin{split}
x & = \arg \, [\, \sin \lambda \, \sin \lambda_c + \cos \lambda \, \cos \lambda_c \, \cos(\phi-\phi_c), \; \cos \lambda \, \sin(\phi-\phi_c) \,], \\
y & = \sin \lambda \, \cos \lambda_c - \cos \lambda \, \sin \lambda_c \, \cos(\phi - \phi_c),
\end{split}
\end{equation}
which effectively performs a spherical coordinate rotation, $(\phi_c,\lambda_c)$ now becoming $(0,0)$. Here the function $\arg(a,b)$ denotes an angle $x$ which has $|\tan x|=|b/a|$ and is also in the same quadrant with the point $(a,b)$. Similar operations can be found in section 2.3 of \cite{calabretta2002}.

This procedure is meant to mitigate the foreshortening effect, so each final image appears as if one is observing directly overhead.  It is important to note that the contours of constant $x$ and $y$ generally do not align with those of longitude and latitude, as illustrated in Figure~\ref{f:cea}.

Furthermore, since the measurement is made in the CCD image, we need to convert the Heliographic coordinate $(\phi,\lambda)$ to the CCD coordinate $(\xi,\eta)$. Assuming the solar radius is $r$ (in pixel), the disk center longitude and latitude is $(\phi_0,b)$, and the image has a $p$-angle (CW rotation on the image needed for the solar north to be $+\eta$), we make use of the following equations
\begin{equation}
\label{eq:imgrot}
\begin{split}
\xi & = \xi' \, \cos p \, - \, \eta' \, \sin p, \\
\eta & = \xi' \, \sin p \, + \, \eta' \, \cos p,
\end{split}
\end{equation}
where
\begin{equation}
\label{eq:sphere2img}
\begin{split}
r & = r_0 \, \cos g \, / \, [\, 1 \, - \, (\sin \lambda \, \sin b + \cos \lambda \, \cos b \, \cos (\phi - \phi_0)) \, \sin g \,], \\
\xi' & = r \, \cos \lambda \, \sin (\phi - \phi_0), \\
\eta' & = r \, [\, \sin \lambda \, \, \cos b \, - \, \cos \lambda \, \sin b \, \cos (\phi - \phi_0) \,].
\end{split}
\end{equation}
This assumes the disk center CCD coordinate is $(0,0)$, and the plate scales in $\xi$ and $\eta$ are identical. The observed solar disk radius is $r_0$, and $g$ is the half angular width of the apparent solar disk. The first equation of equation set (6) accounts for the finite distance between the Sun and SDO.


\subsection{Pipeline Procedure: Sampling}
\label{subsec:sampling}

The remapping module works in the following order. First, we define the final grid in CEA coordinate, centered at the pre-determined location corresponding to $(\phi_c,\lambda_c)$, equally spaced with a grid size of 0.03$^\circ$ in both directions (about 0.5'' at disk center). Second, these sets of $(x,y)$ are converted to $(\phi,\lambda)$, and subsequently to $(\xi,\eta)$. Third, interpolation is performed on the CCD image at $(\xi,\eta)$ to obtain the field value at $(x,y)$, a procedure known as sampling.

A tracking module determines for each set of SHARP its maximum extent in Heliographic longitude ($\Delta \phi$) and latitude ($\Delta \lambda$) during the disk passage. Each cutout is ensured to enclose at least this Heliographic range. The remapping module currently requests a largest possible, constant size in CEA, $\Delta x = \Delta \phi$ and $\Delta y = \Delta \lambda$. A side effect of this choice is that $\Delta x$ and $\Delta y$ may convert to an area slightly larger than what the cutout encloses. A few rows or columns of pixels at the edge might not have the azimuthal 180$^\circ$-ambiguity resolution. They are marked 0 in the \texttt{BIT\_MAP} image for being outside the identified region, and 0 in the \texttt{CONF\_DISAMBIG} image for being not disambiguated.

For HMI vector field, sampling is performed on $B_\xi$, $B_\eta$, and $B_\zeta$ individually, with a sixth-order Wiener interpolation scheme. The same interpolator is used for float quantities, like line-of-sight magnetograms. For \texttt{BIT\_MAP} and \texttt{CONF\_DISAMBIG}, a near-neighbor interpolator is used. Note that data processed before August 2013 used Wiener interpolation for these two, which results in unreasonable values at sharp boundaries (e.g. $-1$ or 35 for \texttt{BIT\_MAP}, 48 or 93 for \texttt{CONF\_DISAMBIG}). They are being reprocessed.

In limited trial cases with visual inspection, weak aliasing patterns were found in $|B|$ images (sampling done on $B_\xi$, etc.) for a grid size of 0.06$^\circ$ using bilinear interpolation. No obvious examples were found with 0.03$^\circ$ resolution using Wiener interpolation.

We nevertheless adopt an oversampling-smoothing scheme to suppress the possible aliasing. We first perform the sampling at a 0.01$^\circ$ resolution. The resultant map is then resized to 0.03$^\circ$, using Gaussian smoothing with $\sigma_x=\sigma_y=0.01^\circ$, truncated at $2\sigma$.


\subsection{Vector Transformation}
\label{subsec:transform}

The field vectors in \texttt{hmi.sharp\_cea\_720s} are presented as $(B_r,B_\theta,B_\phi)$. The base vector $(\hat{\bf{e}}_r,\hat{\bf{e}}_\theta,\hat{\bf{e}}_\phi)$ is that of a Heliocentric spherical coordinate. Note that the subscript $\theta$ here in $B_\theta$ denote \textit{colatitude}, and $\theta=90^\circ-\lambda$. The transformation from the newly sampled $(B_\xi,B_\eta,B_\zeta)$ to $(B_r,B_\theta,B_\phi)$ follows equation (1) of \cite{gary1990}:
\begin{equation}
\label{eq:gary}
\left( \begin{array}{r}
B_r \\
B_\theta \\
B_\phi \end{array} \right) =
\left( \begin{array}{rrr}
k_{11} & k_{12} & k_{13} \\
k_{21} & k_{22} & k_{23} \\
k_{31} & k_{32} & k_{33} \end{array} \right)
\left( \begin{array}{r}
B_\xi \\
B_\eta \\
B_\zeta \end{array} \right),
\end{equation}
where
\begin{equation}
\label{eq:k}
\begin{split}
k_{11} & = \cos \lambda \,[ \sin b \sin p \cos (\phi-\phi_0) + \cos p \sin(\phi-\phi_0)] - \sin \lambda \,[\cos b \sin p], \\
k_{12} & = - \cos \lambda \,[ \sin b \cos p \cos (\phi-\phi_0) - \sin p \sin(\phi-\phi_0)] + \sin \lambda \,[\cos b \cos p], \\
k_{13} & = \cos \lambda \cos b \cos (\phi-\phi_0) + \sin \lambda \sin b, \\
k_{21} & = \sin \lambda \,[ \sin b \sin p \cos (\phi-\phi_0) + \cos p \sin(\phi-\phi_0)] + \cos \lambda \,[\cos b \sin p], \\
k_{22} & = - \sin \lambda \,[ \sin b \cos p \cos (\phi-\phi_0) - \sin p \sin(\phi-\phi_0)] - \cos \lambda \,[\cos b \cos p], \\
k_{23} & = \sin \lambda \cos b \cos (\phi-\phi_0) - \cos \lambda \sin b, \\
k_{31} & = - \sin b \sin p \sin(\phi-\phi_0) + \cos p \cos(\phi-\phi_0), \\
k_{32} & = \sin b \cos p \sin(\phi-\phi_0) + \sin p \cos(\phi-\phi_0), \\
k_{33} & = - \cos b \sin (\phi-\phi_0).
\end{split}
\end{equation}
Here $(\phi,\lambda)$ is the longitude and \textit{latitude} of the pixel, $(\phi_0,b)$ is the longitude and latitude of the disk center, $p$ is the solar $p$-angle. We show in the Appendix the derivation of equation~\eqref{eq:gary} and that $(B_r,B_\theta,B_\phi)$ are identical to $(B_z^h,-B_y^h,B_x^h)$ in \cite{gary1990}.

We note that the sign of $B_\theta$ is positive if the component points toward south. The base vector $\hat{\bf{e}}_\phi$ is parallel to the constant latitude line; $\hat{\bf{e}}_\theta$ parallel to the constant longitude line. As illustrated in Figure~\ref{f:cea}, they are not parallel to $+x$ or $+y$ in the final CEA coordinates away from the patch center. The deviation may be estimated by
\begin{equation}
\label{eq:deviation}
\begin{split}
\cos^{-1} |{\hat{\bf{e}}}_\phi \cdot {\hat{\bf{e}}}_x| & = \cos^{-1} \frac{|\nabla \phi \cdot {\hat{\bf{e}}}_x|}{|\nabla \phi|}
  = \tan^{-1} \frac{|\partial{\phi}/\partial{y}|}{|\partial{\phi}/\partial{x}|} \\
  & = \tan^{-1} \frac{|\sin x \, \sin \lambda_c \, (y\,\sin\lambda_c \, - \, \sqrt{1-y^2} \, \cos x \, \cos \lambda_c )|}
  			    {(1\, - \, y^2)|\cos x \,[(1 \, - \, y^2) \, \cos^2 \lambda_c+y^2 \, \sin^2 \lambda_c]\,-\,y\,\sqrt{1-y^2}\,(1\,+\, \cos^2 x) \, \cos \lambda_c \, \sin \lambda_c|}, \\
			    \\
\cos^{-1} |{\hat{\bf{e}}}_\theta \cdot {\hat{\bf{e}}}_y| & = \cos^{-1} \frac{|\nabla \theta \cdot {\hat{\bf{e}}}_y|}{|\nabla \theta|} 
  = \tan^{-1} \frac{|\partial{\lambda}/\partial{x}|}{|\partial{\lambda}/\partial{y}|} \\
  & = \tan^{-1} \frac{(1-y^2) \sin \lambda_c \sin x}{\sqrt{1-y^2} \cos \lambda_c - y \sin \lambda_c \cos x}.
\end{split}
\end{equation}
For $x=-15^\circ$, $y=-15^\circ$, $\lambda_c=-20^\circ$, i.e. lower-left corner of Figure~\ref{f:cea}, we have $\cos^{-1} |{\hat{\bf{e}}}_\phi \cdot {\hat{\bf{e}}}_x|=6.6^\circ$, $\cos^{-1} |{\hat{\bf{e}}}_\theta \cdot {\hat{\bf{e}}}_y|=5.7^\circ$. The deviation is generally a few degrees at most.

The inversion provides uncertainty estimates on $B$, $\gamma$, and $\psi$, that is $\sigma_B$, $\sigma_\gamma$, and $\sigma_\psi$, respectively. Pairwise correlation functions are also provided, which we convert to the pairwise covariances ${\rm{cov}}_{B\gamma}$, ${\rm{cov}}_{B\psi}$, and ${\rm{cov}}_{\psi\gamma}$. Using equation sets~\eqref{eq:bvec}, \eqref{eq:gary}, and \eqref{eq:k}, we may estimate the uncertainties in $\sigma_{B_r}$, $\sigma_{B_\theta}$, and $\sigma_{B_\phi}$.
\begin{equation}
\label{eq:err1}
\begin{split}
\sigma^2_{B_r} & = \left( \frac{\partial B_r}{\partial B} \right )^2 \sigma^2_B + \left( \frac{\partial B_r}{\partial \gamma} \right )^2 \sigma^2_\gamma + \left( \frac{\partial B_r}{\partial \psi} \right )^2 \sigma^2_\psi + 2 \frac{\partial B_r}{\partial B} \frac{\partial B_r}{\partial \gamma} {\rm{cov}}_{B\gamma} + 2 \frac{\partial B_r}{\partial B} \frac{\partial B_r}{\partial \psi} {\rm{cov}}_{B\psi} + 2 \frac{\partial B_r}{\partial \psi} \frac{\partial B_r}{\partial \gamma} {\rm{cov}}_{\psi \gamma}, \\
\sigma^2_{B_\theta} & = \left( \frac{\partial B_\theta}{\partial B} \right )^2 \sigma^2_B + \left( \frac{\partial B_\theta}{\partial \gamma} \right )^2 \sigma^2_\gamma + \left( \frac{\partial B_\theta}{\partial \psi} \right )^2 \sigma^2_\psi + 2 \frac{\partial B_\theta}{\partial B} \frac{\partial B_\theta}{\partial \gamma} {\rm{cov}}_{B\gamma} + 2 \frac{\partial B_\theta}{\partial B} \frac{\partial B_\theta}{\partial \psi} {\rm{cov}}_{B\psi} + 2 \frac{\partial B_\theta}{\partial \psi} \frac{\partial B_\theta}{\partial \gamma} {\rm{cov}}_{\psi \gamma},\\
\sigma^2_{B_\phi} & = \left( \frac{\partial B_\phi}{\partial B} \right )^2 \sigma^2_B + \left( \frac{\partial B_\phi}{\partial \gamma} \right )^2 \sigma^2_\gamma + \left( \frac{\partial B_\phi}{\partial \psi} \right )^2 \sigma^2_\psi + 2 \frac{\partial B_\phi}{\partial B} \frac{\partial B_\phi}{\partial \gamma} {\rm{cov}}_{B\gamma} + 2 \frac{\partial B_\phi}{\partial B} \frac{\partial B_\phi}{\partial \psi} {\rm{cov}}_{B\psi} + 2 \frac{\partial B_\phi}{\partial \psi} \frac{\partial B_\phi}{\partial \gamma} {\rm{cov}}_{\psi\gamma},
\end{split}
\end{equation}
For simplicity, we use a near-neighbor interpolation to get $(\xi,\eta)$ for the corresponding CEA $(x,y)$. In this way, $\sigma_B$, $\sigma_\gamma$, and $\sigma_\psi$ are directly taken from the inversion at the integer pixel $(\xi,\eta)$, without further expansion of the propagation equations. The partial derivatives are evaluated as
\begin{equation}
\label{eq:err2}
\begin{split}
\frac{\partial B_r}{\partial B} & = -k_{11} \sin \gamma \sin \psi + k_{12} \sin \gamma \cos \psi + k_{13} \cos \gamma, \\
\frac{\partial B_\theta}{\partial B} & = -k_{21} \sin \gamma \sin \psi + k_{22} \sin \gamma \cos \psi + k_{23} \cos \gamma, \\
\frac{\partial B_\phi}{\partial B} & = -k_{31} \sin \gamma \sin \psi + k_{32} \sin \gamma \cos \psi + k_{33} \cos \gamma, \\
\frac{\partial B_r}{\partial \gamma} & = B(-k_{11} \cos \gamma \sin \psi + k_{12} \cos \gamma \cos \psi - k_{13} \sin \gamma), \\
\frac{\partial B_\theta}{\partial \gamma} & = B(-k_{21} \cos \gamma \sin \psi + k_{22} \cos \gamma \cos \psi - k_{23} \sin \gamma), \\
\frac{\partial B_\phi}{\partial \gamma} & = B(-k_{31} \cos \gamma \sin \psi + k_{32} \cos \gamma \cos \psi - k_{33} \sin \gamma), \\
\frac{\partial B_r}{\partial \psi}  & = B(-k_{11} \sin \gamma \cos \psi - k_{12} \sin \gamma \sin \psi), \\
\frac{\partial B_\theta}{\partial \psi}  & = B(-k_{21} \sin \gamma \cos \psi - k_{22} \sin \gamma \sin \psi), \\
\frac{\partial B_\phi}{\partial \psi}  & = B(-k_{31} \sin \gamma \cos \psi - k_{32} \sin \gamma \sin \psi).
\end{split}
\end{equation}


\section{Useful Examples}
\label{sec:example}

This section provides a few examples for vector field data analyzing using \texttt{IDL} and the \texttt{SolarSoft} modules (\texttt{SSW}, \cite{freeland1998}). The dataset used here is the one illustrated in Figure~\ref{f:cea}, and can be downloaded \href{http://jsoc.stanford.edu/jsocwiki/sharp_coord?action=AttachFile&do=get&target=sharp_example.zip}{from this link}. It includes six images (\texttt{field}, \texttt{inclination}, \texttt{azimuth}; \texttt{B\_p}, \texttt{B\_t}, \texttt{B\_r}). The first three are cutout images from \texttt{hmi.sharp\_720s[377][2011.02.16\_18:00]}. The last three are CEA maps from \texttt{hmi.sharp\_cea\_720s[377][2011.02.16\_18:00]}, where \texttt{B\_p} is for $B_\phi$, \texttt{B\_t} is for $B_\theta$.

To reduce the file size, most images are stored in the CFITSIO Rice-compressed format. Most standard FITS viewing software with versions after 2010 also can read this standard compression. They can also be read using the \texttt{SSW} module \texttt{read\_sdo}. Alternatively, one can try the stand-alone fits reader \href{http://hmi.stanford.edu/doc/magnetic/fitsio.pdf}{\texttt{fitsio\_read\_image}}.


\begin{table}[h]
\begin{center}
\caption{Ephemeris keywords for the SHARP series \label{tbl:eph}}
\begin{tabular}{c|c|cc}
\hline
\hline
Keyword (n=1,2) & Meaning & \texttt{hmi.sharp\_720s} & \texttt{hmi.sharp\_cea\_720s} \\
\hline
\texttt{CRPIXn} & ref. pixel & disk center w.r.t. patch lower-left & patch center w.r.t. patch lower-left \\
\texttt{CRVALn} & ref. pixel coord. & Solar $(X,Y)\equiv(0,0)$ & Carrington lon., lat. $(\phi, \lambda)$ \\
\texttt{CRDELTn} & pixel size & $\approx0.5''$ & 0.03$^\circ$ \\
\texttt{CROTA2} & negative $p$-angle & $\approx180^\circ$ & 0$^\circ$ \\
\texttt{IMCRPIXn} & disk center & \multicolumn{2}{c}{pixel coord in 4K image} \\
\texttt{CRLN(T)\_OBS} & disk center & \multicolumn{2}{c}{Carrington lon. (lat.) of disk center} \\
\hline
\end{tabular}
\end{center}
\end{table}


\subsection{Headers: Where Is SHARP on the Sun?}
\label{subsec:header}

The HMI images and their headers (``metadata'') are stored separately and are combined upon ``export''. For an introduction to the full disk image keywords, see the \href{http://jsoc.stanford.edu/doc/keywords/JSOC_Keywords_for_metadata.pdf}{metadata documentation}.

The keywords of the SHARP series follow the WCS convention. The lower-left corner pixel always have the pixel address of (1,1). Some ephemeris keywords are provided in Table~\ref{tbl:eph}.

Examples for the keyword values in \texttt{hmi.sharp\_720s}. It is easy to infer that the lower-left corner of the cutout has an integer pixel address of $(685,2354)$ in the 4K image, assuming the lower-left is $(1,1)$.
\begin{alltt}
\colorbox{grey}{IDL> read_sdo, `hmi.sharp_720s.377.20110216_180000_TAI.field.fits', index, data}
\colorbox{grey}{IDL> print, index.crpix1, index.crpix2, index.crval1, index.crval2, \$}
\colorbox{grey}{IDL>        index.cdelt1, index.cdelt2, index.crota2, format=`(2f10.3,5f9.3)'}
\colorbox{grey}{  1354.933  -306.263    0.000    0.000    0.504    0.504  180.083}
\colorbox{grey}{IDL> print, index.imcrpix1, index.imcrpix2  ;ad hoc, disk center pixel in 4K image}
\colorbox{grey}{       2038.9334       2046.7371}
\end{alltt}
For \texttt{hmi.sharp\_cea\_720s}.
\begin{alltt}
\colorbox{grey}{IDL> read_sdo, `hmi.sharp_cea_720s.377.20110216_180000_TAI.Br.fits', index, data}
\colorbox{grey}{IDL> print, index.crpix1, index.crpix2, index.crval1, index.crval2, index.cdelt1, \$}
\colorbox{grey}{IDL>        index.cdelt2, index.crln_obs, index.crlt_obs, format=`(2f10.3,6f9.3)'}
\colorbox{grey}{   372.500   189.000  395.183  -21.077    0.030    0.030  359.790   -6.877}
\end{alltt}


\subsection{The Handy \texttt{SSW} Modules}
\label{subsec:trans}

The existing WCS-related modules in \texttt{SolarSoft} are very handy for coordinate conversion and are in good agreement with the HMI pipeline results. The WCS information is retrieved from the FITS header and used for the transformation. For the cutout series, one can do the following.
\begin{alltt}
\colorbox{grey}{IDL> read_sdo, `hmi.sharp_720s.377.20110216_180000_TAI.field.fits', index, data}
\colorbox{grey}{IDL> wcs = fitshead2wcs(index)  ;extract WCS info}
\colorbox{grey}{IDL> coord = wcs_get_coord(wcs)  ;get coordinate values}
\end{alltt}

To convert the current coordinate $(\xi,\eta)$ to $(X_\odot,Y_\odot)$, i.e. the westward and northward distances to the disk center, one uses the following command.
\begin{alltt}
\colorbox{grey}{IDL> wcs_convert_from_coord, wcs, coord, `HPC', solar_x, solar_y, /arcseconds}
\end{alltt}
This results in two maps named \texttt{solar\_x} and \texttt{solar\_y}, which contain the Solar X and Y coordinate for each pixel in the cutout image. Here, `HPC' stands for Helioprojective-Cartesian.

Similarly, one uses the following o convert $(\xi,\eta)$ to longitude and latitude $(\phi,\lambda)$. `HG' stands for Heliographic. Note it assumes the disk center longitude $\phi_0$ is 0$^\circ$ (Stonyhurst Heliographic).
\begin{alltt}
\colorbox{grey}{IDL> wcs_convert_from_coord, wcs, coord, `HG', phi, theta}
\end{alltt}

The same command sequence applies to the CEA series. See the documentation of \texttt{wcs\_convert\_from\_coord} in \texttt{SSW} for details.

One may, of course, implement equation sets~\eqref{eq:cea} to~\eqref{eq:sphere2img} and their inverses to perform the conversion. Those are the ones used in the HMI pipeline.

The difference between the \texttt{SSW} version and the HMI pipeline is generally small, less than 0.1\% of the grid size. The exception is when the absolute value of the solar radius is used, e.g. during conversion from $(\phi,\lambda)$ to $(\xi,\eta)$. The default value for \texttt{SSW} (set in the module \texttt{wcs\_rsun}) is 6.955d8, whereas the HMI pipeline uses 6.96d8. Once the value in \texttt{SSW} is changed to 6.96d8, the resultant $(\xi,\eta)$ agrees with the HMI pipeline within a few percent. The rest of the difference comes from the different treatment of the finite distance between the Sun and SDO.


\subsection{Constructing Radial Field Cutout}
\label{subsec:radial}

The HMI pipeline infers field strength $B$, inclination $\gamma$, and azimuth $\psi$, which are provided in the cutout series. The longitude and latitude are readily available from the previous example. Equation sets~\eqref{eq:bvec} and~\eqref{eq:gary} give the method to obtain $B_r$ from these images.  


\section{Outlook}
\label{sec:outlook}

Future plans include a web tool that allows the coordinate system conversion upon request. A variety of coordinates for remapping and vector transformation will be available.


%
\subsection*{Acknowledgements}
The original remapping code \texttt{cartography.c} is developed by Rick Bogart. The original vector data processing code is written by Yang Liu. The usage of several \texttt{SSW} modules on HMI data was first suggested by Harry Warren.

%
\subsection*{Appendix}

We show here that the field components $(B_r,B_\theta,B_\phi)$ in our Heliocentric spherical coordinate are identical to the ``Heliographic components'' $(B_z^h,-B_y^h,B_x^h)$ in \cite{gary1990}.

We make use of a Heliocentric Cartesian coordinate where the Solar rotational axis is $+Z$, the line pointing from Sun center to the longitude and latitude point $(0,0)$ is $+X$. Conversion between the two requires a rotation with respect to $+\zeta$ by $p$, followed by a rotation with respect to the new $+\xi$ by $b$, and a permutation of the axes. This means
\begin{equation}
\label{eq:p2c}
\left( \begin{array}{r}
B_X \\
B_Y \\
B_Z \end{array} \right) =
\left( \begin{array}{rrr}
0 & 0 & 1 \\
1 & 0 & 0 \\
0 & 1 & 0 \end{array} \right)
\left( \begin{array}{rrr}
1 & 0 & 0 \\
0 & \cos b & \sin b \\
0 & -\sin b & \cos b \end{array} \right)
\left( \begin{array}{rrr}
\cos p & \sin p & 0 \\
-\sin p & \cos p & 0 \\
0 & 0 & 1 \end{array} \right)
\left( \begin{array}{r}
B_\xi \\
B_\eta \\
B_\zeta \end{array} \right).
\end{equation}

Further, converting from $(B_X,B_Y,B_Z)$ to $(B_r,B_\theta,B_\phi)$ needs
\begin{equation}
\label{eq:c2r}
\left( \begin{array}{r}
B_r \\
B_\theta \\
B_\phi \end{array} \right) =
\left( \begin{array}{rrr}
\cos \lambda \cos(\phi-\phi_0) & \cos \lambda \sin(\phi-\phi_0) & \sin \lambda \\
\sin \lambda \cos(\phi-\phi_0) & \sin \lambda \sin(\phi-\phi_0) & -\cos \lambda \\
-\sin(\phi-\phi_0) & \cos(\phi-\phi_0) & 0 \end{array} \right)
\left( \begin{array}{r}
B_X \\
B_Y \\
B_Z \end{array} \right).
\end{equation}
Again, here $\lambda$ is the latitude. The subscript $\theta$ in $B_\theta$ refers to \textit{colatitude}, and is positive when pointing south.

Equation~\eqref{eq:gary} can then be derived by combining \eqref{eq:p2c} and \eqref{eq:c2r}. Compare the coefficients $k_{ij}$ in \eqref{eq:k} with $a_{ij}$ in equation (1) of \cite{gary1990}, it is easy to verify that
\begin{equation}
\label{eq:a}
\left( \begin{array}{r}
B_r \\
B_\theta \\
B_\phi \end{array} \right) =
\left( \begin{array}{rrr}
a_{31} & a_{32} & a_{33} \\
-a_{21} & -a_{22} & -a_{23} \\
a_{11} & a_{12} & a_{13} \end{array} \right)
\left( \begin{array}{r}
B_\xi \\
B_\eta \\
B_\zeta \end{array} \right) =
\left( \begin{array}{r}
B_z^h \\
-B_y^h \\
B_x^h \end{array} \right).
\end{equation}

%
%
 \bibliographystyle{spr-mp-sola-cnd} 
 \bibliography{coord}  

\end{document}